\documentclass[preprint,12pt]{elsarticle}
\usepackage[cp1251]{inputenc}

\usepackage{amssymb}

\usepackage{booktabs} 
\usepackage{color}
\usepackage[cp1251]{inputenc}
\usepackage[english]{babel}
\usepackage{amsfonts,amssymb,amsmath}
\usepackage{amsmath,graphicx}
\usepackage{graphicx,pifont} 
\usepackage{epsfig}
\usepackage{parallel}
\usepackage{esdiff}

\begin{document}

\begin{frontmatter}



\title{Field-theoretical form of the molecular dynamics method in condensed matter physics: relativistic quantum statistical thermodynamics}


\author{A. Yu. Zakharov}

\affiliation{organization={Dept. General and Experimental Physics,\\ Yaroslav-the-Wise Novgorod State University}, 
            addressline={41, B. Sanct-Petersburgskaya}, \\
            city={Veliky Novgorod},
            postcode={173003}, 
            country={Russian Federation}}
\ead{anatoly.zakharov@novsu.ru}
\begin{abstract}
A method for quantitatively describing the relativistic dynamics of a classical system of interacting atoms is proposed and developed within the concept of an auxiliary field. For atoms at rest, this field is equivalent to a given static interatomic potential, and under dynamic conditions, it represents a relativistic scalar field. The auxiliary dynamic field created by atoms is a superposition of elementary fields of the Klein–Gordon type, whose mass parameters are uniquely determined by singular points of the Fourier transform of the corresponding static interatomic potential. The system of interacting atoms is represented as the union of two ideal subsystems, one of which is a system of free atoms (an ideal gas), and the other is a system of free massive oscillators (a thermostat). Thermal equilibrium between these subsystems is ensured by the interaction between the atoms and the auxiliary fields. It is shown that the divergence of the classical partition function of the auxiliary field (analogous to the ultraviolet catastrophe) is eliminated within the framework of quantum theory. The quantum partition function of the auxiliary field has been calculated exactly.

\end{abstract}


\end{frontmatter}

\section{Introduction}

Classical Gibbs statistical mechanics aims to provide a microscopic foundation of thermodynamics~\cite{Gibbs1902},
basing on two fundamental \textit{postulates}:

\begin{enumerate}
	\item The microscopic dynamics of a system of particles (atoms) is subject to classical Newtonian mechanics, presented in Hamiltonian form.
	\item A state of thermodynamic equilibrium of a macroscopic many-body system exists and, depending on external conditions (fixed energy E, temperature T, or chemical potential $\mu$), obeys axiomatically introduced corresponding (microcanonical, canonical, or grand canonical) distributions of statistical mechanics. In terms of modern probability theory, these distributions are probability measures in the phase space of the many-body system.
\end{enumerate}

Direct implementation of the Gibbs method comes down to solving two problems.
\begin{itemize}

	\item 
	The problem of describing interactions between the structural units of matter (atoms). It is generally assumed that this interaction can be represented in terms of instantaneous interatomic potentials, which at best can exist only in the non-relativistic approximation, i.e., for particles at rest. However, even in this approximation, systematic methods for finding interatomic potentials are unknown. Therefore, ``model potentials'' that exhibit ``correct'' asymptotic behavior are typically used. ~\cite{Kaplan, Kamberaj}.

	\item 
	The problem of exact calculation the partition function of many-body systems with a given model interatomic potential. Despite enormous efforts over many years, the number of model interatomic potentials for which exact solutions have been found remains extremely limited~\cite{Baxter, Grosse, McCoy, Sutherland}. Significantly, not a single three-dimensional model has yet been solved.
	
\end{itemize}  

Besides the insurmountable \textbf{computational} difficulties, there are at least two unsolved \textbf{fundamental problems} in statistical mechanics.

\begin{enumerate}
	\item 
	\textbf{Physical problem}: the zeroth law of thermodynamics, postulated in both thermodynamics and statistical mechanics, is fundamentally incompatible with classical Newtonian mechanics~\cite{Uhlenbeck}. The essence of the matter is that within the framework of classical mechanics of a system of interacting particles, there is no mechanism leading to the irreversibility of the system's dynamics. Therefore, a consistent microscopic foundation of thermodynamics within the framework of classical mechanics is fundamentally impossible~\cite{Zakharov2025}.

	\item \textbf{Mathematical problem}: the concept of probability based on Lebesgue measure theory~\cite{Kolmogorov}, has an inevitable  mathematical ambiguity due to Ulam's theorem~\cite{Ulam1930,Ulam1960}, according to which there is no countably additive measure $\mu(A)$ defined for \textit{all} subsets $A$ of a set $E$ of power continuum~($\aleph_{1} $).
In statistical mechanics, the set of elementary events~$E$ is the set of all points in the phase space~$\Gamma$ of the system~-- this is a set of power~($\aleph_{1} $). Therefore, different measures in the phase space of a system of particles correspond to different (generally speaking, nonequivalent) probability models~\cite{Khrennikov2009, Khrennikov2016, Khrennikov2023}. In particular, the question of the equivalence conditions of the basic probability measures (i.e., the microcanonical, canonical, and grand canonical ensembles) of statistical mechanics remains open~\cite{Ellis2000, Campa, Zhang2023}.

\end{enumerate}

One of the options for a microscopic explanation and justification of thermodynamics is the relativistic field model of a system of interacting particles (atoms). Within this model, the system consists of two subsystems, one of which are the atoms themselves, and the other of which is the field, which mediates the interactions between the atoms.

The fundamental difference between a relativistic field system of interacting particles and a non-relativistic one is that the set of degrees of freedom of the "atoms + field" system consists of a finite number of atomic degrees of freedom and an infinite set of field degrees of freedom. Essentially, the system of atoms is immersed in the variable field they create, which acts as a hidden, irremovable thermostat. Therefore, thermodynamic equilibrium in a system of interacting atoms is an equilibrium between the atoms, on one side, and the field (thermostat) they create, on the other. The energy of interatomic interactions is the energy of the field that mediates the interactions between the atoms. 

Thus, in non-relativistic dynamics in general and classical statistical mechanics in particular, the infinite set of field degrees of freedom of a system is completely ignored, and only a finite number of atomic degrees of freedom are considered. The evolution of the system as a whole depends on the initial conditions of both the atoms and the field. Therefore, specifying the initial conditions of the atoms alone is insufficient to unambiguously determine the dynamics of both the system as a whole and the dynamics of the atoms in particular.

In turn, due to the finite propagation velocity of the field, the instantaneous configuration of the field generated by atoms depends on the prehistory of the atomic dynamics. This leads to the phenomenon of hereditary behavior, including the effect of lag in interatomic interactions. The works ~\cite{Zakharov2016, Zakharov2019} have shown that retardation in interactions between atoms is a real physical mechanism leading to the irreversibility of the dynamics of an atomic system. Thus, there is no need to appeal probabilistic hypotheses to explain the phenomenon of irreversibility from a microscopic perspective.

\section{The concept of an auxiliary field. Lagrangian picture}

The concept of an auxiliary field was proposed and developed in the works~\cite{Zakharov2022, Zakharov2022-2, Zakharov2023}. By definition, an auxiliary field~$\varphi\left(\mathbf{r}, t\right)$ is a function that satisfies the following conditions:
\begin{enumerate}
\item for atoms at rest~$\varphi\left(\mathbf{r}, t\right)$ is equivalent to a given central interatomic potential $v\left( r\right) $, $r =\left| \mathbf{r}\right| $;
\item for atoms in motion~$\varphi\left(\mathbf{r}, t\right)$ is a covariant relativistic field. 
\end{enumerate}
For the class of interatomic potentials that can be represented as a Fourier integral
\begin{equation}\label{Four2}
	v\left( r\right) = \int\, \dfrac{d \mathbf{k} }{\left(2\pi \right)^{3} }\, \tilde{v}\left(k^{2} \right) \, e^{i \mathbf{k r}},
\end{equation}
the equation for the free auxiliary field has the following form:
\begin{equation}\label{varphi-0,t}
	\left( \tilde{v}\left( -\square\right)\right)^{-1}  \varphi\left( \mathbf{r},t\right) = 0,
\end{equation}	
where $\square = \dfrac{\partial^{2}}{\partial x^{2}} + \dfrac{\partial^{2}}{\partial y^{2}} + \dfrac{\partial^{2}}{\partial z^{2}} - \dfrac{1}{c^{2}}\,\dfrac{\partial^{2}}{\partial t^{2}} $~is the d'Alembert operator.

To each singular point (pole) $k_{s}$ of the function~$\tilde{v}\left(k^{2} \right) $ on the complex plane of the variable~$k$ there corresponds an \textit{elementary auxiliary field}~$\varphi_{s} \left(\mathbf{r}, t \right) $ satisfying an equation of the Klein-Gordon type with, generally speaking, a complex parameter $\mu_{s}=i k_{s}$:
\begin{equation}\label{partial-s}
	\left( \square - \mu_{s}^{2}\right)^{\nu_{s}}\varphi_{s}\left(\mathbf{r}, t \right) = 0,
\end{equation} 
where~$\nu_{s}$~is the pole multiplicity.

The complete auxiliary field $\varphi\left(\mathbf{r}, t \right)$ is a superposition (linear combination) of the elementary auxiliary fields~$\varphi_{s}\left(\mathbf{r}, t \right)$
\begin{equation}\label{superpos}
\varphi\left( \mathbf{r}, t\right) = \sum_{s} C_{s}\, \varphi_{s}\left( \mathbf{r}, t\right).
\end{equation}
Thus, within the framework of relativistic theory, the auxiliary field that ensures interatomic interactions is completely characterized by the set of singular points of the function $\tilde{v}\left( k^{2}\right) $, the existence of which is guaranteed by Liouville's theorem~\cite{Whittaker}.

We restrict ourselves to the case where the multiplicities of all poles of the function~$\tilde{v}\left( k^{2}\right) $ are equal to unity~$\nu_{s}=1$ and the parameters~$\mu_{s}$ are real.
The complete system of equations describing the dynamics of atoms and the auxiliary field they create follows from the variational principle for the action functional~\cite{Zakharov2022, Zakharov2022-2}:
\begin{equation}\label{Action}
	\begin{array}{r}
		{\displaystyle 	S=-\sum\limits_{a} m c\int \ ds_{a}-\sum_{s=1}^{n} \sum_{a}\frac{\gamma_{s}}{c}\int \varphi_{s}(x_{a}) \ ds_{a} }\\
		{\displaystyle + \sum_{s=1}^{n}\frac{\varkappa_{s}}{2c} \int   d^{4}x\,   \left( \partial_{\nu} \varphi_{s}(x)\, \partial^{\nu}\! \varphi_{s}(x) - \mu_{s}^2\varphi_{s}^2(x) \right),}
	\end{array}
\end{equation}
where $m$ and $s_{a}$~are the mass and world line of the $a$-th atom, respectively, $\gamma_{s}$ and $\varkappa_{s}$~are constants. 

Note that the action functional is based on the assumption that atoms do not undergo destruction over time, which in reality can only occur at sufficiently high speeds. Therefore, we will assume that the characteristic velocities of atoms, $v$, are much less than the speed of light, $c$:
\begin{equation}\label{v<<c}
	v \ll c.
\end{equation}
As a result, the actual expression for the action functional of the system ``atoms + auxiliary field'' has the form:
\begin{equation}\label{Action2}
	\begin{array}{r}
		{\displaystyle 	S = \int  dt\left\lbrace \sum\limits_{a} \frac{m  \dot{\mathbf{R}}_{a}^{2}\left( t\right) }{2} -\sum_{s=1}^{n} \gamma_{s} \int\limits_{\left( V\right) } d\mathbf{r} \left( \sum_{a} \delta \left(\mathbf{r - R}_{a}\left(t \right)  \right) \right)  \varphi_{s}(\mathbf{r})  \right. }\\
		{\displaystyle + \sum_{s=1}^{n}\frac{\varkappa_{s}}{2} \int\limits_{\left( V\right) }  d\mathbf{r} \, \left[   \left. \left(  \frac{\partial \varphi_{s}(\mathbf{r}, t)}{c\, \partial t}\right)^{2}   -  \left(\nabla \varphi_{s}(\mathbf{r}, t) \right)^{2} -  \mu_{s}^2\varphi_{s}^2(\mathbf{r}, t)\right] \right\rbrace ,}
	\end{array}
\end{equation}
where $V$~is the volume of the system.
The expression contained in the curly brackets of this formula is the Lagrangian of the system:
\begin{equation}\label{Lagrange}
	\begin{array}{r}
		{\displaystyle 	L \left(\mathbf{R}_{a}\left( t\right),   \dot{\mathbf{R}}_{a}\left( t\right); \left\lbrace  \varphi_{s}(\mathbf{r}, t) \right\rbrace,  \left\lbrace  \dot {\varphi}_{s}(\mathbf{r}, t) \right\rbrace, \left\lbrace  \nabla \varphi_{s}(\mathbf{r}, t) \right\rbrace \right) }\\
		{\displaystyle =  \sum\limits_{a} \frac{m \dot{\mathbf{R}}_{a}^{2}\left( t\right) }{2} -\sum_{s=1}^{n} \gamma_{s} \int\limits_{\left( V\right) } d\mathbf{r} \left( \sum_{a} \delta \left(\mathbf{r - R}_{a}\left(t \right)  \right) \right)  \varphi_{s}(\mathbf{r}, t)  }\\
		{\displaystyle + \sum_{s=1}^{n}\frac{\varkappa_{s}}{2} \int\limits_{\left( V\right) }  d\mathbf{r} \, \left[      \left(  \frac{\partial \varphi_{s}(\mathbf{r}, t)}{c\, \partial t}\right)^{2}   -  \left(\nabla \varphi_{s}(\mathbf{r}, t) \right)^{2} -  \mu_{s}^2\varphi_{s}^2(\mathbf{r}, t)\right]  .}
	\end{array}
\end{equation}
From here follow both the equations of the dynamics of atoms 
\begin{equation}\label{atoms}
 m \ddot{\mathbf{R}}_{a}\left( t\right) + \sum_{s}\, \gamma_{s}\nabla \varphi_{s}\left(\mathbf{R}_{a}\left( t\right)  \right) = 0,
\end{equation}
and the equations of the evolution of the elementary auxiliary fields of the system:
\begin{equation}\label{el-field}
 \left(\square -\mu_{s}^{2} \right) \varphi_{s}\left( \mathbf{r}, t\right) = \frac{\gamma_{s}}{\varkappa_{s}} \sum_{a} \delta\left( \mathbf{r - R}_{a}\left( t\right)  \right).
\end{equation}

The evolution equations of each of the elementary auxiliary fields admit an exact analytical solution~$\varphi_{s}\left( \mathbf{r}, t\right)$, which depends on the trajectories of all atoms~$\mathbf{R}_{a}\left(t' \right) $ for all $t'\in \left(-\infty < t' < t \right) $.
Substituting the functions~$\varphi_{s}\left(\mathbf{r}, t \right) $ into the equations~\eqref{atoms} leads to the elimination of field variables and the appearance of a closed system of equations for~$\mathbf{R}_{a}\left(t\right) $.
It is clear that the solution of this system of equations~$\mathbf{R}_{a}\left( t\right) $ depends not only on the initial conditions for all atoms, but also on the entire history of all atoms at all previous times $t'<t $.

\section{Hamiltonian of a system ``atoms and an auxiliary field''}

Let us pass from the elementary auxiliary fields $\varphi_{s}\left(\mathbf{r}, t \right) $ to their Fourier components $\tilde{v}_{s}\left(\mathbf{k}, t \right)$~\cite{Haar}:
\begin{equation}\label{Four-0}
	\varphi_{s}\left( \mathbf{r}, t\right) = \sum_{\mathbf{k}} \tilde{\varphi}_{s}\left(\mathbf{k}, t \right) e^{i \mathbf{k r}}, \quad \tilde{\varphi}_{s}\left(\mathbf{k}, t \right) = \frac{1}{V}\int\limits_{\left( V\right) } \varphi_{s}\left(\mathbf{r}, t \right) e^{-i \mathbf{k r}}\, d\mathbf{r}.
\end{equation}

Note that the Fourier components~$\varphi_{s} \left(\mathbf{k}, t \right) $ are complex-valued and have the following properties
\begin{equation}\label{var-phi}
\tilde{\varphi}_{s}\left(-\mathbf{k}, t \right) = \tilde{\varphi}^{*}_{s}\left(\mathbf{k}, t \right).
\end{equation}
To each complex-valued function~$\varphi_{s}\left(\mathbf{k}, t \right) $ we assign a pair of real field variables  ${\psi}_{s}\left(\mathbf{k}, t \right)$ and ${\chi}_{s}\left(\mathbf{k}, t \right)$:
\begin{equation}\label{var--phi}
	\tilde{\varphi}_{s}\left(\pm \mathbf{k}, t \right) = {\psi}_{s}\left(\mathbf{k}, t \right) \pm i{\chi}_{s}\left(\mathbf{k}, t \right),
\end{equation}
that have the following properties
\begin{equation}\label{pm}
	\left\lbrace 
	\begin{array}{l}
	{\displaystyle {\psi}_{s}\left(-\mathbf{k}, t \right) = {\psi}_{s}\left(\mathbf{k}, t \right),}\\
	{\displaystyle {\chi}_{s}\left(-\mathbf{k}, t \right) = -{\chi}_{s}\left(\mathbf{k}, t \right)}.
	\end{array}
	\right. 
\end{equation}

Let's transform the Lagrangian~\eqref{Lagrange} to new variables~$\mathbf{R}_{a}\left( t\right),   \dot{\mathbf{R}}_{a}\left( t\right)$;    $ \psi_{s}\left(\mathbf{k}, t \right), \dot{\psi}_{s}\left(\mathbf{k}, t \right)$;  $  \chi_{s}\left(\mathbf{k}, t \right), \dot{\chi}_{s}\left(\mathbf{k}, t \right)  $: 
\begin{equation}\label{Lagrange1}
	\begin{array}{c}
		{\displaystyle 	L \left(\mathbf{R}_{a}\left( t\right),   \dot{\mathbf{R}}_{a}\left( t\right);   {\psi}_{s}(\mathbf{k}, t),   \dot{\psi}_{s}\left(\mathbf{k}, t\right); {\chi}_{s}(\mathbf{k}, t),   \dot{\chi}_{s}\left(\mathbf{k}, t\right) \right)  =  \sum\limits_{a} \frac{m \dot{\mathbf{R}}_{a}^{2}\left( t\right) }{2}  }\\
		{\displaystyle    -  \sum_{s=1}^{n} \gamma_{s} \sum_{a, \mathbf{k}}\,\left[  \psi_{s} \left(\mathbf{k}, t \right)\, \cos\left(\mathbf{k R}_{a}\left(t \right)  \right)  - \chi_{s}\left( \mathbf{k}, t\right) \sin\left( \mathbf{k R}_{a}\left(t \right) \right)   \right]  }\\ 
				{\displaystyle  + \sum_{s=1}^{n}\frac{V\varkappa_{s}}{2c^{2}} 
		\sum_{\mathbf{k}} \left[ \dot{\psi}_{s}^{2}\left(\mathbf{k}, t\right)  - c^{2}\left(k^{2} + \mu^{2}_{s} \right) {\psi}_{s}^{2}\left(\mathbf{k}, t \right) \right] }\\
			{\displaystyle  + \sum_{s=1}^{n}\frac{V\varkappa_{s}}{2c^{2}} 
			\sum_{\mathbf{k}} \left[ \dot{\chi}_{s}^{2}\left(\mathbf{k}, t\right)  - c^{2}\left(k^{2} + \mu^{2}_{s} \right) {\chi}_{s}^{2}\left(\mathbf{k}, t \right) \right] .}\\
		\end{array}
\end{equation}

Substituting the Lagrangian~\eqref{Lagrange1} into the Lagrange equations leads to the following equations of motion of atoms and evolution of elementary auxiliary fields:
\begin{equation}\label{eq-at-f2}
	\left\lbrace 
	\begin{array}{l}
		{\displaystyle m \ddot{\mathbf{R}}_{a} \left( t\right) - \sum_{s=1}^{n}\gamma_{s} \sum_{\mathbf{k}} \mathbf{k}\left[ \psi_{s} \left(\mathbf{k}, t \right)\, \sin\left(\mathbf{k R}_{a}\left(t \right)  \right) + \chi_{s}\left( \mathbf{k}, t\right) \cos\left( \mathbf{k R}_{a}\left(t \right) \right) \right] =0;  }\\
		{\displaystyle  \ddot{\psi} _{s}\left(\mathbf{k}, t \right) +c^{2} \left(k^{2} + \mu_{s}^{2} \right) {\psi} _{s}\left(\mathbf{k}, t \right) + \frac{\gamma_{s} c^{2}}{V\varkappa_{s}} \sum_{a}\cos \left(\mathbf{k R}_{a} \left( t\right) \right) =0;}\\
		{\displaystyle \ddot{\chi} _{s}\left(\mathbf{k}, t \right) +c^{2} \left(k^{2} + \mu_{s}^{2} \right) {\chi} _{s}\left(\mathbf{k}, t \right) - \frac{\gamma_{s} c^{2}}{V\varkappa_{s}} \sum_{a}\sin \left(\mathbf{k R}_{a} \left( t\right) \right) =0 . }
	\end{array}
	\right. 
\end{equation} 
The first of these equations describes the dynamics of atoms in an evolving auxiliary field, and the second and third ones describe the evolution of elementary auxiliary fields created by moving atoms.


Let us move from the Lagrangian description of the dynamics of a system of interacting atoms to the Hamiltonian picture. We define the momenta of the atoms $\mathbf{P}_{a}\left( t \right) $ and of the elementary auxiliary fields $p_{s}\left( \mathbf{k}, t\right),\ \mathfrak{p}_{s} \left( \mathbf{k}, t\right) $ by the relations:
\begin{equation}\label{p_a}
	\left\lbrace 
	\begin{array}{l}
		{\displaystyle \mathbf{P}_{a} \left( t \right) = \frac{ \partial L}{\partial \dot{\mathbf{R}}_{a} \left( t \right) } = m \dot{\mathbf{R}}_{a}\left( t\right); }\\ 
		{\displaystyle p_{s}\left( \mathbf{k}, t\right) = \frac{\partial L}{\partial  \dot{\psi}_{s}\left(\mathbf{k}, t \right)}  = \frac{V\varkappa_{s}}{c^{2}} \, \dot{\psi}_{s} \left(\mathbf{k},t \right) ;}\\
		{\displaystyle \mathfrak{p}_{s}\left( \mathbf{k}, t\right) = \frac{\partial L}{\partial  \dot{\chi}_{s}\left(\mathbf{k}, t \right)}  = \frac{V\varkappa_{s}}{c^{2}} \, \dot{\chi}_{s} \left(\mathbf{k},t \right) .}
	\end{array}
	\right. 
\end{equation}

We perform the Legendre transformation and find the Hamiltonian of a system consisting of atoms and auxiliary fields
\begin{equation}\label{Hamilton1}
	\begin{array}{c}
		{\displaystyle 	H \left(\mathbf{R}_{a}\left( t\right),   {\mathbf{P}}_{a}\left( t\right);   {\psi}_{s}(\mathbf{k}, t),  {p}_{s}(\mathbf{k}, t) ; \chi_{s}\left(\mathbf{k}, t \right),  \mathfrak{p}_{s}\left(\mathbf{k}, t \right) \right) = \sum\limits_{a} \frac{\mathbf{P}_{a}^{2}\left( t\right) }{2 m}} \\
		{\displaystyle 
			 +\sum_{s=1}^{n} \gamma_{s} \sum_{a, \mathbf{k}}\,\left[  \psi_{s} \left(\mathbf{k}, t \right)\, \cos\left(\mathbf{k R}_{a}\left(t \right)  \right)  - \chi_{s}\left( \mathbf{k}, t\right) \sin\left( \mathbf{k R}_{a}\left(t \right) \right)   \right] }\\
		{\displaystyle + \sum_{s=1}^{n} \sum_{\mathbf{k}} \left[   \frac{c^{2}}{V\varkappa_{s}} \frac{p_{s}^{2} \left( \mathbf{k}, t\right)}{2}  +   V\varkappa_{s} \left(k^{2} + \mu^{2}_{s} \right) \frac{{\psi}_{s}^{2}\left(\mathbf{k}, t \right)}{2}  \right] }\\
		{\displaystyle + \sum_{s=1}^{n} \sum_{\mathbf{k}} \left[   \frac{c^{2}}{V\varkappa_{s}} \frac{\mathfrak{p}_{s}^{2} \left( \mathbf{k}, t\right)}{2}  +   {V\varkappa_{s}} \left(k^{2} + \mu^{2}_{s} \right) \frac{{\chi}_{s}^{2}\left(\mathbf{k}, t \right)}{2}  \right] .}
	\end{array}
\end{equation}

The first term on the right-hand side of this formula is the Hamiltonian of the subsystem of free atoms, the third and fourth terms are the Hamiltonian of the subsystem of free massive (due to the parameters $\mu_{s}$) oscillators with field coordinates ${\psi}_{s}\left(\mathbf{k}, t \right)$, ${\chi}_{s}\left(\mathbf{k}, t \right)$ and momenta $p_{s} \left(\mathbf{k}, t\right)$, $\mathfrak{p}_{s} \left(\mathbf{k}, t\right)$, respec\-tively, and the second term is the interaction between atoms and the auxiliary field, which implements the energy exchange between both subsystems. Ther\-mo\-dynamic equilibrium in the system as a whole will be interpreted as equilibrium between these subsystems. 

Thus, we will represent a system of interacting atoms as a union of two interpenetrating, inseparable substances:
\begin{enumerate}
	\item an ideal gas consisting of atoms;
	\item an ideal gas consisting of a set of oscillators characterized by mass parameters~$\mu_{s}$.
\end{enumerate}
In this representation, the auxiliary field acts as an inseparable heat reservoir with respect to the subsystem of atoms.

\section{Classical partition function of the auxiliary field}

Let's extract from the Hamiltonian~\eqref{Hamilton1} the part corresponding to the free auxiliary field:

 \begin{equation}\label{H-f}
 H_{f} = \sum_{s}\, H_{f_{s}}\left(p_{s}\left(\mathbf{k}, t \right), \psi_{s}\left(\mathbf{k}, t \right); \mathfrak{p}_{s}\left(\mathbf{k}, t \right), \chi_{s}\left(\mathbf{k}, t \right) \right),
 \end{equation}
where
\begin{equation}\label{H-fs}
\begin{array}{r}
	{\displaystyle  H_{f_{s}} \left(p_{s}\left(\mathbf{k}, t \right), \psi_{s}\left(\mathbf{k}, t \right); \mathfrak{p}_{s}\left(\mathbf{k}, t \right), \chi_{s}\left(\mathbf{k}, t \right) \right)}\\
	{\displaystyle = \sum_{\mathbf{k}}\left[   \frac{c^{2}}{V\varkappa_{s}} \frac{p_{s}^{2} \left( \mathbf{k}, t\right)}{2}  +   V\varkappa_{s} \left(k^{2} + \mu^{2}_{s} \right) \frac{{\psi}_{s}^{2}\left(\mathbf{k}, t \right)}{2}  \right]  }\\
	{\displaystyle  +  \sum_{\mathbf{k}}\left[   \frac{c^{2}}{V\varkappa_{s}} \frac{\mathfrak{p}_{s}^{2} \left( \mathbf{k}, t\right)}{2}  +   V\varkappa_{s} \left(k^{2} + \mu^{2}_{s} \right) \frac{{\chi}_{s}^{2}\left(\mathbf{k}, t \right)}{2}  \right] }
\end{array}
\end{equation}
is the Hamiltonian of the $s$-th elementary auxiliary field~$\varphi_{s}\left(\mathbf{r}, t\right) $.

The classical canonical partition function of the $s$-th elementary auxiliary field is
\begin{equation}\label{Z-s-0}
\begin{array}{r}
{\displaystyle Z_{s} = \idotsint \left(\prod_{\mathbf{k}} dp_{s}\left(\mathbf{k} \right) d\psi_{s}\left( \mathbf{k}\right) \, d\mathfrak{p}_{s}\left( \mathbf{k}\right) d\chi_{s}\left( \mathbf{k}\right) \right)  }\\
{\displaystyle \times \exp\left[  {-\beta H_{f_{s}} \left(p_{s}\left(\mathbf{k} \right), \psi_{s}\left(\mathbf{k} \right); \mathfrak{p}_{s}\left(\mathbf{k} \right), \chi_{s}\left(\mathbf{k} \right) \right) }\right]  },
\end{array}	
\end{equation}
where $\beta= 1/k_{B}T$, $k_{B}$~is the Boltzmann constant. 
This integral is Gaussian and can be calculated exactly.
The result is as follows:
\begin{equation}\label{Z-s}
Z_{s} = \prod_{\mathbf{k}}\left[\frac{2\pi}{\beta \omega_{s}\left(\mathbf{k} \right) } \right]^{2}, 
\end{equation}
where 
\begin{equation}\label{omega-k}
\omega_{s}\left(\mathbf{k} \right) = c \sqrt{\mathbf{k}^{2}+\mu_{s}^{2}}.
\end{equation}

From here we find the partition function~$Z$ of the complete composite auxiliary field
\begin{equation}\label{Z-full}
	Z = \prod_{s} Z_{s} = \prod_{s, \mathbf{k}} \left[\frac{2\pi}{\beta \omega_{s}\left(\mathbf{k} \right) } \right]^{2} 
\end{equation} 
and the average energy of the auxiliary field: 
%
%
%
\begin{equation}\label{<E>}
\left\langle E\right\rangle = - \frac{\partial	}{\partial \beta}\left( \ln Z\right) = 2 \sum_{s, \mathbf{k}} k_{B}T.
\end{equation}
This sum over $s, \mathbf{k}$ contains infinitely many identical terms, so the average energy~$\left\langle E\right\rangle$ is infinite:
\begin{equation}\label{infty}
	\left\langle E\right\rangle = \infty.
\end{equation}
 
The resulting infinity of auxiliary field energy is analogous to the paradox associated with the radiation of a heated body, known since the late 19th century as the ultraviolet catastrophe. In 1900, this paradox was resolved by Planck through the introduction of the quantum hypothesis.

\section{Quantum statistics of elementary auxiliary fields}


Let us make the transition from the classical description of these elementary auxiliary fields~$\psi_{s}\left(\mathbf{k} \right) $, $\chi_{s}\left(\mathbf{k} \right) $ to the quantum one. We assume that these fields are bosonic. Since in their dispersion law~\eqref{omega-k} the mass parameters~$\mu_{s}$ are nonzero, we will call them \textit{massive phonons}. Their energy spectrum is characterized by a continuous vector~$\mathbf{k}$ and nonnegative integer quantum occupation numbers $n_{s}=0, 1, 2, 3, \ldots$:
\begin{equation}\label{quant}
E_{n}^{s}\left(\mathbf{k} \right)  = \left( n_{s}+\frac{1}{2}\right) \hbar\omega_{s}\left( \mathbf{k}\right). 
\end{equation}   
From here we find the connection between the infinitesimal changes of $k$ and $E_{n}\left(k \right) $:
\begin{equation}\label{dE-dk}
	dE_{n}^{s}\left( k\right) =\left(n_{s} + \frac{1}{2} \right) \frac{c\hbar k}{\sqrt{k^{2} + \mu_{s}^{2}}} dk.
\end{equation}
%
%

The number of states $N_{s}$ of the auxiliary field $\varphi_{s}\left(\mathbf{ k }\right) $ with  $\left|\mathbf{k} \right| \leq k$  is
\begin{equation}\label{Ns}
N_{s} = \frac{V}{\left( 2\pi\right)^{3}} \, \frac{4}{3}\pi k^{3}
\end{equation}  
%
This implies that the number of states~$dN_{s}\left( k\right) $ in the layer between~$k$ and $k+dk$ 
\begin{equation}\label{dNs}
dN_{s}\left( k\right)  = \frac{V}{\left(2\pi \right)^{3} }\, 4\pi k^{2}\,dk.
\end{equation}

The dispersion law of massive oscillators (phonons) is as follows
\begin{equation}\label{Es}
E_{s}\left( \mathbf{k}\right) =\hbar\, \omega_{s} \left( \mathbf{k}\right) = \hbar c \sqrt{\mu_{s}^{2} + k^{2}}.
\end{equation}
%
Let us express $dE_{s}(\mathbf{k})$ in terms of $dk$:
\begin{equation}\label{dEs-k}
d\left(\hbar \omega_{s}\left(\mathbf{k} \right)  \right) = \hbar c \frac{k}{\sqrt{\mu_{s}^{2} + k^{2}}}\, dk.
\end{equation}
%
The partition function of an individual harmonic~$\omega_{s}\left( \mathbf{ k} \right) $ has the known form
\begin{equation}\label{Z-s(omega)}
Z_{s}\left(\mathbf{k} \right) = \sum\limits_{n=0}^{\infty} e^{-\beta \hbar \omega_{s}\left(\mathbf{k} \right)\left[ n + \frac{1}{2}\right]  } = \frac{e^{-\frac{\beta\hbar \omega_{s}\left( \mathbf{k} \right) }{2}}}{1-e^{-\beta\hbar\omega_{s}\left( \mathbf{k}\right) }}.
\end{equation}
%
From here we find the average energy of this harmonic
\begin{equation}\label{E-harm}
E_{s}\left(\mathbf{k} \right) = - \frac{\partial \ln Z_{s}\left( \mathbf{k}\right) }{\partial \beta} = \frac{\hbar \omega_{s}\left(\mathbf{k} \right)}{2} + \frac{\hbar \omega_{s}\left(\mathbf{k} \right)}{e^{\beta\hbar \omega_{s}\left(\mathbf{k} \right)} -1}.
\end{equation}
Since at $T=\frac{1}{\beta}\to 0$ the second term on the right side of this formula tends to zero, the first term represents the energy of the ground state, which we will take as the initial point of reference for the energy:
\begin{equation}\label{U-s}
U_{s} = \int\limits_{0}^{\infty} \frac{\hbar \omega_{s}\left(\mathbf{k} \right)}{e^{\beta\hbar \omega_{s}\left(\mathbf{k} \right)} -1} dN_{s}\left( k\right) = 
\frac{V}{\left(2\pi \right)^{3} } \int\limits_{0}^{\infty} \frac{k^{2}}{2\pi^{2}} \frac{\hbar c \sqrt{\mu_{s{\tiny }}^{2} + k^{2} }}{e^{\beta\hbar c \sqrt{\mu_{s}^{2} + k^{2} }} -1}dk . 
\end{equation} 
%
%
Let's introduce dimensionless variables~
\begin{equation}\label{Q-Ms}
Q = \beta \hbar c k, \quad \tau_{s} = \left( \beta \hbar c \mu_{s}\right) ^{-1} = {T}/{T_{s}}, 
\end{equation}
where $T_{s}=\hbar c \mu_{s}$~ is the relativistic quantum energy characteristic of the elementary auxiliary field~$\varphi_{s}\left(\mathbf{r}, t \right) $, $T$~ is the absolute temperature in energy units ($k_{B}=1$), $\tau_{s}={T}/{T_{s}}$~ is the dimensionless temperature of the auxiliary field.

In the new variables, the expression for~$U_{s}$ has the form
\begin{equation}\label{dim-Us}
	U_{s}  =  
	\frac{V}{\left(2\pi \right)^{4} \beta^{4} \left( \hbar c \right)^{3} } \int\limits_{0}^{\infty} \frac{Q^{2}}{\pi} \frac{ \sqrt{\tau_{s}^{-2} + Q^{2} }}{e^{ \sqrt{\tau_{s}^{-2} + Q^{2} }} -1}dQ .
\end{equation}

Note that the factor before the integral in this formula is common to all elementary auxiliary fields and is proportional to~$T^{4}$ , while the ``individual'' integral
\begin{equation}\label{J(Ms)}
	J\left(\tau_{s} \right) = \int\limits_{0}^{\infty} \frac{Q^{2}}{\pi} \frac{ \sqrt{\tau_{s}^{-2} + Q^{2} }}{e^{ \sqrt{\tau_{s}^{-2} + Q^{2} }} -1}dQ
\end{equation}
depends only on the dimensionless temperature~$\tau$.

The function~$J\left( \tau\right) $ is a monotonically increasing function of~$\tau$. Its limit values are as follows:
\begin{equation}\label{lims-J}
\lim\limits_{\tau \to +0} J\left( \tau\right) =0; \quad \lim\limits_{\tau \to \infty} J\left( \tau\right) = \pi^{3}/15\approx 2.067.
\end{equation}
Note that when~$\tau\to 0$, not only the function~$J\left(\tau \right) $, but also all of its derivatives, vanish.
At $\tau \lesssim 0.07$ the function $J\left(\tau \right) $ is almost identically equal to zero, then quickly reaches saturation at~$\tau \approx 3$, and then tends to~$\pi^3/15$.

The graph of the function~$J\left(\tau \right) $ is shown in Fig. 1.
\begin{figure}
	\centering
	\includegraphics[width=1.0\linewidth]{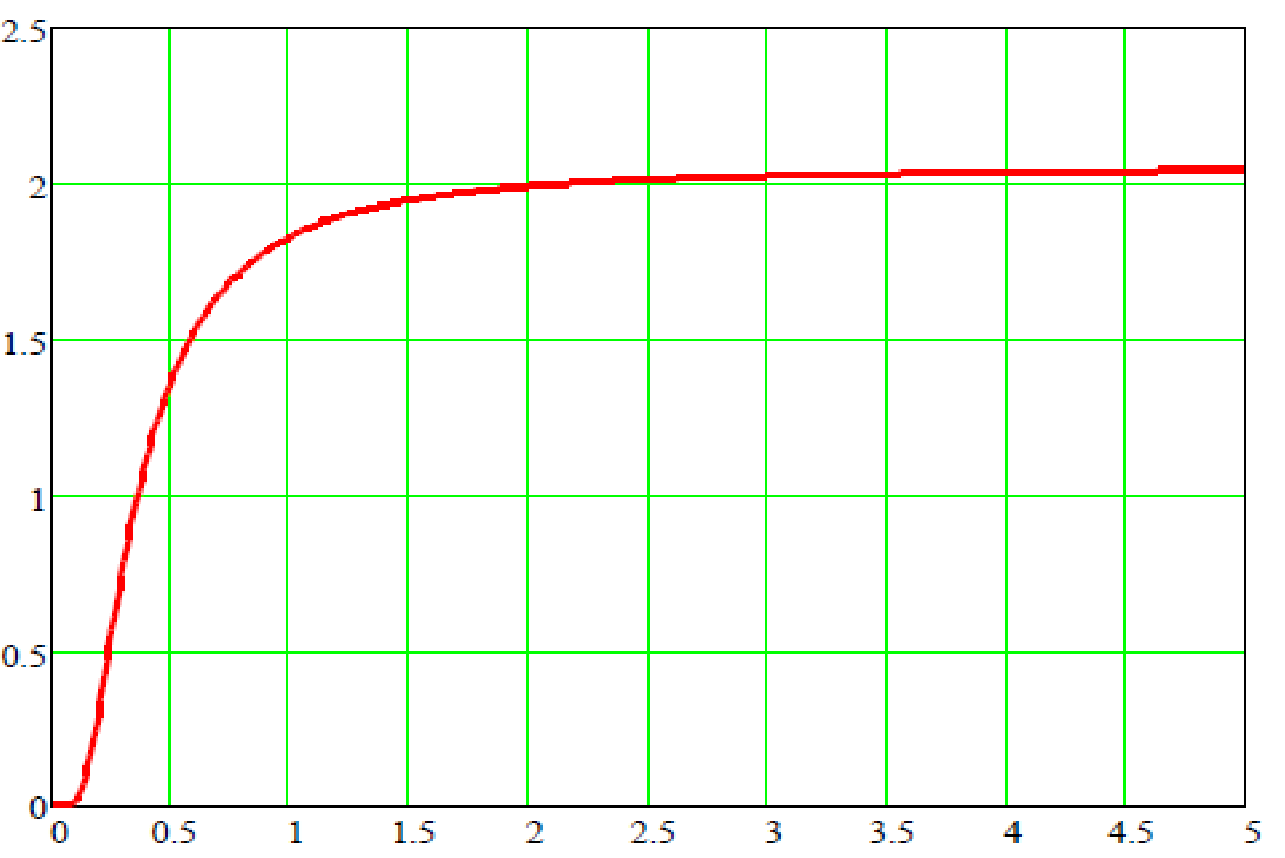}
	\caption{Dependence of $J\left(\tau \right)$ on the dimensionless temperature~$\tau$.}
	\label{fig:1-line}
\end{figure}

\section{Energy of the total auxiliary field}

The total auxiliary field of a system of interacting atoms~$\varphi\left( \mathbf{r}, t\right) $ is a linear combination of elementary auxiliary fields~$\varphi_{s}\left( \mathbf{r}, t\right) $, each of which has its own characteristic temperature~$T_{s}=\hbar c \mu_{s}$.

The graphs of the functions $J(tau_s)$ of all elementary auxiliary fields at their individual dimensionless temperatures $\tau$ are identical. However, it is interesting to compare the graphs of these functions in terms of a common reference temperature $T_{0}$. The different elementary fields $\varphi_s\left(\mathbf{r}, t\right)$ and $\varphi_{s'}\left(\mathbf{r}, t\right)$ differ from each other only in their mass parameters $\mu_s$ and $\mu_{s'}$, and hence in their characteristic temperatures $T_{s}$ and $T_{s'}$, respectively.
Therefore, to calculate the curves~$J\left( \tau\right) $ of different elementary fields in terms of the common reference temperature~$T_{0}$, one should make the substitution in the formula~\eqref{J(Ms)}
\begin{equation}\label{a-tau}
	\tau \Rightarrow \frac{\mu_{0}}{\mu_{s}}\tau,
\end{equation}
where $\mu_{0}$ and $\mu_{s}$ are the mass parameters of the reference and studied elementary fields, respectively.


The results of calculations of the curves~$J\left( \tau\right) $ of several elementary auxiliary fields relative to the common reference temperature $T_{0}=\hbar c \mu_{0}$ are presented in Fig. 2.
\begin{figure}
	\centering
	\includegraphics[width=0.9\linewidth]{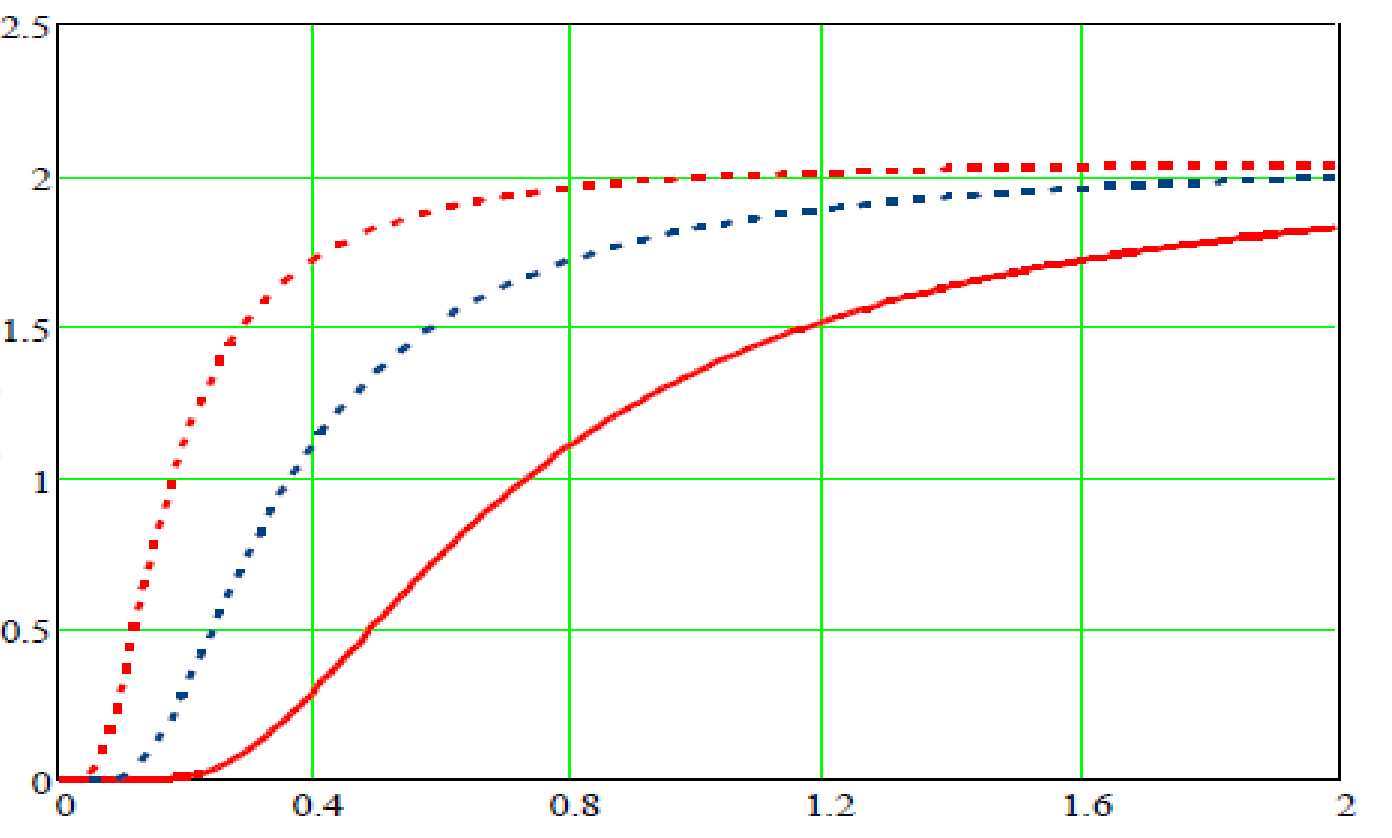}
	\caption{Temperature dependences of the functions $J_{s}\left(\tau \right)$ of elementary fields with masses~$\mu_{0}$ (blue dotted line), $\mu_{1}=2\mu_{0}$ (red solid line), $\mu_{2}=0.5\mu_{0}$ (red dotted line)}
	\label{fig:3-lines}
\end{figure}

The total energy of the auxiliary field~$\varphi\left(\mathbf{r}, t \right) $ is the sum of the contributions of the elementary fields~$\varphi_{s}\left(\mathbf{r}, t \right) $. At low temperatures, the main contribution to the energy comes from fields with minimal masses. As the temperature increases, contributions from more massive fields arise and increase.

\section{Conclusion}

The main results of this work are as follows.
\begin{enumerate}
	\item A Hamiltonian description of the relativistic dynamics for a system of interacting atoms is proposed within the framework of the concept of a covariant auxiliary field, taking into account both atomic and field degrees of freedom.
	\item It is shown that the divergence of the classical partition function of the auxiliary field (analogous to the ultraviolet catastrophe) is eliminated by transitioning to quantum theory. An exact expression for the quantum partition function of the auxiliary field is obtained.
	\item 
	An interpretation of the process of establishing thermal equilibrium in a system of interacting atoms as a process of energy exchange between the auxiliary field and the atoms is proposed.
\end{enumerate}

\section*{Acknowledgements}
I am grateful to Ya.I. Granovsky, M.A. Zakharov, and V.V. Zubkov for fruitful discussions.


\begin{thebibliography}{00}


\bibitem{Gibbs1902} J. W. Gibbs. \textit{Elementary Principles in Statistical Mechanics, developed with especial reference to the rational foundation of thermodynamics}. New York: Charles Scribner’s Sons, 1902. XVIII+244~pp.


\bibitem{Kaplan}  I. G. Kaplan.  \textit{Intermolecular Interactions: Physical Picture, Computational Methods and Model Potentials}. Chichester: Wiley, 2006. 375~p.





\bibitem{Kamberaj}  H. Kamberaj. \textit{Molecular Dynamics Simulations in Statistical Physics: Theory and Applications}. Cham: Springer, 2020. 470~p.


\bibitem{Baxter}  ({R.J.~Baxter}. \textit{Exactly Solved Models in Statistical Mechanics}. London~e.a.: Academic Press. 1982).



\bibitem{Grosse} {H. Grosse.} \textit{Models in Statistical Physics and Quantum Field Theory}. Berlin~e.a.: Springer, 1989. 151~p.


	\bibitem{McCoy} {B. M. McCoy, T. T. Wu.} \textit{The Two-Dimensional Ising Model}. Cambridge, Massachusetts: Harvard University Press, 1973. 418~p.


	\bibitem{Sutherland} {B. Sutherland.} \textit{Beautiful Models. 70~Years of Exactly Solved Quantum Many-Body Problems}. Singapore: World Scientific, 2004. XV+381~p.


\bibitem{Uhlenbeck} 
 G.~E. Uhlenbeck, G. W.~Ford. \textit{Lectures in Statistical Mechanics}. Providence: AMS, 1963. X+171~p. 


\bibitem{Zakharov2025} A. Yu. Zakharov. On the microscopic origin of thermodynamics and kinetics. Status and prospects. Physics Letters A, 2025. 


\bibitem{Kolmogorov}  
A.~Kolmogoroff.  \textit{Grundbegriffe der Wahrscheinlichkeitsrechnung}. Berlin: Springer, 1933. 62~S. 


\bibitem{Ulam1930} S. M. Ulam. Zur Ma{\ss}theorie in der allgemeinen Mengenlehre. \textit{Fundamenta Mathematicae}. {1930}. Vol.{16}, No.1. Pp.140--150. DOI: 10.4064/fm-16-1-140-150 

\bibitem{Ulam1960} S. M. Ulam. \textit{Problems in Modern Mathematics}. New York: Interscience Publisher, 1960. 168~p.


\bibitem{Khrennikov2009} A. Yu. Khrennikov. \textit{Interpretations of probability}. Berlin:  Walter de Gruyter, 2009. 217~p.

\bibitem{Khrennikov2016} A. Yu. Khrennikov. {\textit{Probability and Randomness: Quantum Versus Classical}}. London: Imperial College Press, 2016. 282~p.


\bibitem{Khrennikov2023} A. Yu. Khrennikov. Bild Conception of Scientific Theory Structuring in Classical and Quantum Physics. 
\textit{Entropy}. 2023; \textbf{25}(11), 1565. DOI: 10.3390/e25111565

\bibitem{Ellis2000} R. S. Ellis, K. Haven, B. Turkington. Large Deviation Principles and Complete Equivalence and Nonequivalence Results for Pure and Mixed Ensembles. \textit{Journal of Statistical Physics}. 2000. \textbf{101}(5-6), 999–1064. DOI:10.1023/A:1026446225804

\bibitem{Campa} A.Campa, T. Dauxois, S. Ruffo. Statistical mechanics and dynamics of solvable models with long-range interactions. \textit{Physics Reports}. 2009. \textbf{480}(3-6), 57-159. DOI: 10.1016/j.physrep.2009.07.001

\bibitem{Zhang2023}  Qi Zhang, D. Garlaschelli. Ensemble nonequivalence and Bose–Einstein condensation in weighted networks. Chaos, Solitons \& Fractals. 2023. \textbf{172}, 113546. DOI: 10.1016/j.chaos.2023.113546


\bibitem{Zakharov2016} { A. Yu. Zakharov,  M. A. Zakharov.} Classical many-body sys\-tems with retarded interactions: dynamical irreversibility. 
\textit{Physics Letters A}. 2016. Vol.{380}, No.3. Pp.365--369. DOI:10.1016/j.physleta.2015.10.056


\bibitem{Zakharov2019}  A. Yu. Zakharov. On physical principles and mathematical mechanisms of the phenomenon of irreversibility: \textit{Physica A: Statistical Mechanics and its Applications}. 2019. Vol.{525}. Pp.1289--1295. https://doi.org/10.1016/j.physa.2019.04.047



\bibitem{Zakharov2022} A. Yu. Zakharov, V. V. Zubkov. Field-Theoretical Representation of Interactions between Particles: Classical Relativistic Probability-Free Kinetic Theory. \textit{Universe}, {2022}; \textbf{8}, 281, pp.1--11. \verb! DOI: 10.3390/universe8050281!.


\bibitem{Zakharov2022-2} A. Yu. Zakharov. Field Form of the Dynamics of Classical Many- and Few-Body Systems: From Microscopic Dynamics to Kinetics, Thermodynamics and Synergetics. \textit{Quantum Reports}, 2022; \textbf{4}(4), 533--543. DOI: 10.3390/quantum4040038.


\bibitem{Zakharov2023}  A. Yu. Zakharov, M. A. Zakharov. Relativistic model of interatomic interactions in condensed systems. \textit{Condensed Matter and Interphases}, 2023; \textbf{25}(4), 494--504. DOI: 10.17308/kcmf.2023.25/11480.

\bibitem{Whittaker}  E. T. Whittaker, G. N. Watson. \textit{A course of modern analysis}. Cambridge: Cambridge University Press, 1927. IV+608~p.  

\bibitem{Haar} D. ter Haar. \textit{Elements of Hamiltonian mechanics}. Oxford: Pergamon Press, 1971. 191~pp.



\end{thebibliography}
\end{document}